\documentclass[draft]{agujournal2019}
\usepackage{url} 
\usepackage{lineno}
\usepackage[inline]{trackchanges} 
\usepackage{soul}
\usepackage{amsmath,amssymb}
\draftfalse

\journalname{Geophysical Research Letters}

\begin{document}

\title{Occurrence of Non-Stationarity at Earth's Quasi-Perpendicular Bow Shock}

\authors{Ajay Lotekar\affil{1}\thanks{deceased},  Yuri V. Khotyaintsev\affil{1,2}, Daniel B. Graham\affil{1}, Andrew Dimmock\affil{1}, Andreas Johlander\affil{1,3} and Ahmad Lalti\affil{4} }

\affiliation{1}{Swedish Institute of Space Physics, Uppsala, Sweden}
\affiliation{2}{Space and Plasma Physics, Department of Physics and Astronomy, Uppsala University, Uppsala, Sweden}
\affiliation{3}{Swedish Defence Research Agency, Stockholm, Sweden}
\affiliation{4}{Northumbria University, Newcastle upon Tyne, UK}

\correspondingauthor{Yuri Khotyaintsev}{yuri@irfu.se}

\begin{keypoints}
\item We investigate the statistical occurrence of ion-scale ripples at Earth’s quasi-perpendicular bow shock.
\item Ripples are found for about 2/3 of the shocks examined, and thus the ripples are common at the Earth’s bow shock.
\item The occurrence of ripples is highest for $M_A \gtrsim 6$ and has no dependence on the shock geometry ($\theta_{Bn}$).
\end{keypoints}

%
%

%
%


\begin{abstract}

Collisionless shocks can exhibit non-stationary behavior even under steady upstream conditions, forming a complex transition region. Ion phase-space holes, linked to shock self-reformation and surface ripples, are a signature of this non-stationarity. We statistically analyze their occurrence using 521 crossings of Earth's quasi-perpendicular bow shock. Phase-space holes appear in 65\% of cases, though the actual rate may be higher as the holes may not be resolved during fast shock crossings. The occurrence rate peaks at 70\% for shocks with Alfv\'en Mach numbers $M_A>7$.
These findings suggest that Earth's quasi-perpendicular bow shock is predominantly non-stationary.

\end{abstract}

\section*{Plain Language Summary}
Collisionless shock waves are a fundamental process in solar and astrophysical plasmas. Earth's bow shock, formed by the solar wind hitting the magnetosphere, is one such shock. Weak shocks are stabilized by resistivity, but strong, high-Mach-number shocks require ion reflection to persist. This reflection makes the shock unsteady over time. We investigate this unsteadiness using ion holes -- features in the ion distribution linked to surface ripples. High-resolution data from NASA's Magnetospheric Multiscale mission reveal ion holes in most bow shock crossings, indicating that Earth's bow shock is generally in a non-stationary state.

\section{Introduction}

Plasma shocks are ubiquitous in the universe, occurring around planets \cite{russell1985planetary}, at the outer periphery of the heliosphere \cite{jokipii_heliospheric_2013}, in the solar wind, and in supernova remnants \cite{bell2013cosmic, helder_observational_2012}. In the heliosphere, particle collisions are so infrequent that their effects on the shock are negligible; hence, these shocks are termed collisionless. Collisionless shocks efficiently convert kinetic energy into thermal energy through interactions between particles and electromagnetic fields. The structure of collisionless shocks is highly complex, with internal structures spanning multiple scales and being intricately coupled to shock parameters. Consequently, linking the magnetic structure of the shock front to energy dissipation processes has been a focus of extensive research for decades, yet many questions remain unanswered.

Collisionless shocks are characterized by parameters such as the Alfv\'en Mach number $(M_A)$, plasma beta $(\beta)$ and geometry $(\theta_{Bn})$, which is the angle between the upstream magnetic field and the shock normal direction ($\mathbf{\hat n}$). In this context, quasi-parallel shocks have $\theta_{Bn}<45^\circ$ and quasi-perpendicular shocks have $\theta_{Bn}>45^\circ$. The shock geometry plays a pivotal role in how plasma is processed, and this dictates both the structure of the shock itself and the plasma dynamics upstream and downstream. In this report, we consider only quasi-perpendicular shocks.

For quasi-perpendicular shocks, a fraction of the incident ions can be reflected and gyrate immediately upstream, gaining energy from the solar wind convective electric field before passing downstream. The typical characteristics of a quasi-perpendicular shock are the foot formed by reflected ions, ramp, and overshoot regions. In addition, the reflected ions cause a high anisotropy in the velocity distribution \cite{sckopke1983}, which can subsequently drive waves and instabilities such as whistler waves in the foot \cite{lalti2022whistler} and ion cyclotron waves downstream of the shock. At supercritical shocks, which is typically the case at the Earth's bow shock, these processes (i.e., ion reflection, whistler waves) play a fundamental role in energy dissipation. 

For stable upstream conditions, shocks can develop non-stationary behavior. At high $M_A$ the shock can undergo reformation, which is a cyclic large-scale restructuring of the shock front \cite{russell_overshoots_1982, lembege1987self,Lembege1989form, Winterhalter1088Obser,  krasnoselskikh2002nonstationarity, dimmock2019direct, madanian_dynamics_2020}. Development of the Alfv\'en ion cyclotron (AIC) instability can generate ion-scale ripples on the shock surface \cite{Winske1988Mag}. Both ripples and reformation lead to large variations in the ramp and ion reflection \cite{Burgess2006shock, lobzin2007,Johlander2018, khotyaintsev2024ionreflectionrippledperpendicular}. Using 3D simulations \citeA{burgess2016} have shown that the ion-scale non-stationary shock structure is related to the coupling of the field-parallel propagating fluctuations with the reflected ions. \citeA{Moullard2006Ripples} presented such observations from the Cluster spacecraft, which were attributed to ripples. Afterwards, \citeA{Johlander2016rippled} reported observations of surface ripples at the quasi-perpendicular shock using MMS observations. In that study, ripples were confirmed using multi-spacecraft observations and high-cadence plasma measurements. The ripple period was comparable to the local ion gyroperiod. \citeA{Johlander2018} found that the dispersive properties of the ripples were similar to those of Alfvén ion cyclotron waves, that ripples have linear polarization, propagate in the coplanarity plane at the local Alfvén speed, and have a wavelength somewhat larger than the upstream ion inertial length. 

Most shock ripple investigations have been based on simulations and detailed case studies, so it is unknown how common non-stationarity occurs at the bow shock and how non-stationarity depends on the shock conditions. We, therefore, conduct a statistical study of shock ripples at the quasi-perpendicular bow shock that links their occurrence and properties to shock parameters and upstream conditions. 

\section{Methodology} \label{sec:method}
\begin{figure}[!h]
 \center
\includegraphics[width=0.5\textwidth]{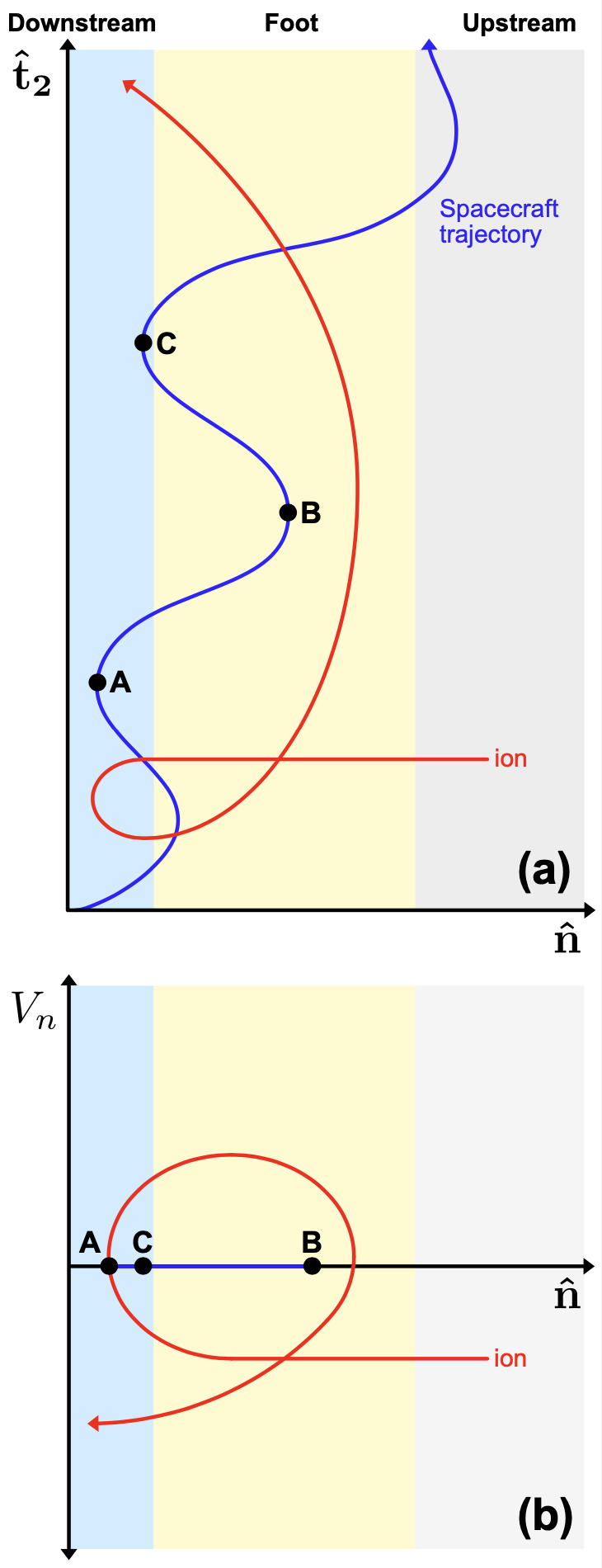}
\caption{Schematic of ion reflection by a perpendicular shock. The ion trajectory is shown in red, and the spacecraft trajectory for a non-stationary (rippled) shock is shown in blue.}
\label{fig:schematic}
\end{figure}

To investigate the presence of non-stationarity at the bow shock, we exploit the reported observations by \cite{Johlander2016rippled, Johlander2018} that revealed shock ripples were accompanied by phase-space holes (PSHs) in the 1D reduced ion velocity distribution along the shock normal direction, $\mathbf{\hat n}$. We illustrate the relation between the ripples and the PSHs in Figure~\ref{fig:schematic}, which shows a trajectory of an ion reflected by a perpendicular shock. We use a $\mathbf{\hat{n}}$, $\mathbf{\hat{t}}_1$, $\mathbf{\hat{t}}_2$ system, where $\mathbf{\hat{t}}_2 = \mathbf{\hat{n}} \times \mathbf{B}_u/|\mathbf{\hat{n}} \times \mathbf{B}_u|$ ($B_u$ is the upstream B) and $\mathbf{\hat{t}}_1 = \mathbf{\hat{t}}_2 \times \mathbf{\hat{n}}$. An ion approaches the shock from upstream, encounters the ramp (foot-downstream boundary) where it is reflected by the normal electric field $E_n$ and cyclotron turning in the stronger downstream magnetic field (panel a). After reflection, the ion gyrates in the upstream field and is accelerated by the upstream convection electric field $\mathbf{E_u}$ (aligned with $\mathbf{\hat{t}_2}$) and returns back to the ramp. This portion of the orbit forms the shock foot. Due to the returning ion's higher energy compared to the initial encounter of the ramp, the ion overcomes the cross-shock potential and is transmitted downstream. In phase space ($V_n$-${\hat{n}}$, Figure \ref{fig:schematic}b), the ion trajectory will correspond to a loop, starting from $V_n<0$ in the upstream, changing to $V_n>0$ after the reflection, and then changing back to $V_n<0$ after the gyration in the upstream field. 
 
The reduced VDF in the $V_n$-${\hat{n}}$ plane exhibits the same pattern as the single particle orbit and forms a PSH in the middle of the loop due to the low upstream thermal speed compared with the bulk speed. A spacecraft can cross such a structure in two ways, both of which will result in an observation of a PSH. Option (I) is the spacecraft crossing a stationary shock. Alternative option (II) is a crossing of a non-stationary shock, where the ripples propagating along the shock plane result in oscillatory motion of the spacecraft relative to the shock, as illustrated by the blue trajectory in Figure \ref{fig:schematic}a. We consider a part of the orbit A-B-C, where the spacecraft starts at A, close to the reflection point $V_n \sim 0$,  moves to B (the foot), where both incident $V_n<0$ and reflected $V_n>0$ ions are present, but no ions with $V_n \sim 0$, and finally moves to C, i.e., returns back to the reflection point ($V_n \sim 0$). A critical difference between options I and II is that option II will result in a relatively symmetric PSH (due to the out-in motion). It can result in multiple PSHs being observed if the overall shock speed is slow compared to the in-plane speed of the ripple. We note that such multiple PSHs produced in option II are not to be confused with the holes generated in the non-linear stage of two-stream instability \cite{omura1996_electron_beam}, as these are apparent holes produced by the motion in and out of the shock foot. Option I will result in a more asymmetric PSH, which is more "open" on the upstream side. This is because of a larger spread in phase space during the upstream gyration compared to the initial reflection at the ramp. 

\begin{figure}[!h]
 \center
\includegraphics[width=0.75\textwidth]{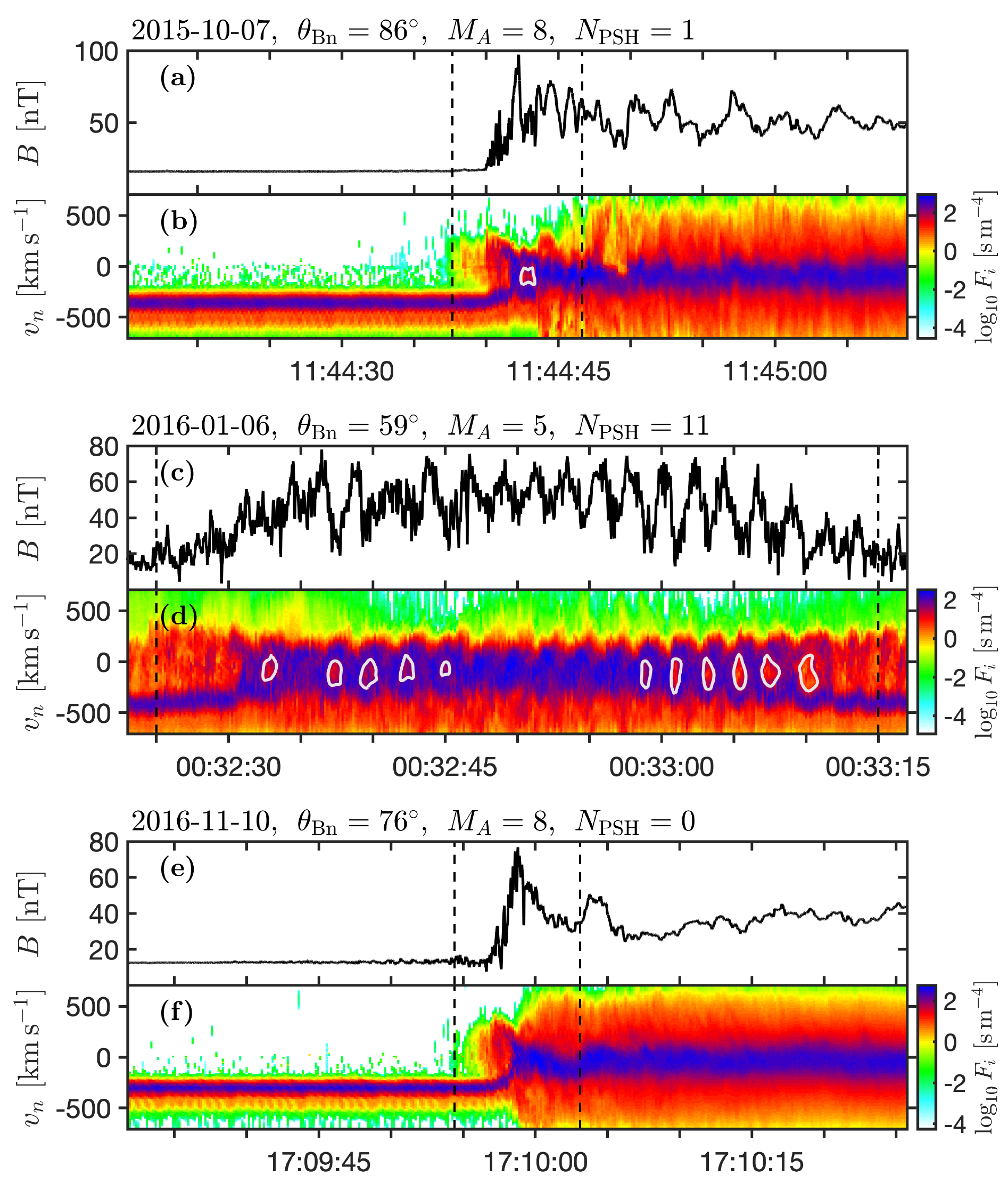}
\caption{Identification of PSHs for three different shocks that exhibit varying numbers of PSHs. The white contours indicate the identified PSHs. The data used are magnetic field at a 128 Hz provided by the fluxgate magnetometers (FGM) \cite{russell_magnetospheric_2016} and 1D reduced ion VDFs calculated from 3D distributions provided by the fast plasma investigators dual ion spectrometer FPI-DIS at 150 ms cadence \cite{pollock_fast_2016}.}
\label{fig:PSH_ID}
\end{figure}

Figure \ref{fig:PSH_ID} shows three shocks of varying PSH numbers. Each sub-figure shows the date, parameters, and number of PSHs for each shock crossing. Figures \ref{fig:PSH_ID}a and \ref{fig:PSH_ID}b show a shock crossing where a single ion hole is identified (white contour line in Figure \ref{fig:PSH_ID}b) from \citeA{Johlander2016rippled}. Figures \ref{fig:PSH_ID}c and \ref{fig:PSH_ID}d show a slow shock crossing from \citeA{Johlander2018}, where multiple symmetric PSHs are observed, consistent with option II. The final example in Figures \ref{fig:PSH_ID}e and \ref{fig:PSH_ID}f contains no PSHs, despite observed reflected ions. Such an event can correspond to a fast shock crossing, where the reflection point is encountered only once (type I). The reflected ions exhibit an ''open" VDF structure produced by option I. 

We now focus on the symmetric PSHs of type II, i.e., the hole referring to the drop in phase space radially bounded by the reflected ions and the reflection point. Observation of multiple such PSHs is evidence for the shock being non-stationary (ripples, reformation). However, observing a single PSH of type II suggests a fast crossing of the reflection point and, thus, non-stationarity of the shock.

We developed an automated procedure to identify PSHs and the number of PSHs observed in each shock. We adopted a contour approach since PSHs can be deemed local minima in phase space. The procedure is as follows:
\begin{enumerate}
    \item Reduce the 3D velocity distribution along $\mathbf{\hat n}$.
    \item Apply a Gaussian filter (fixed filter parameters for all shocks) to remove the impact from small-scale (smaller than the hole) variations and prevent over-identification.
    \item Manually select the crossing time ($\sim$foot), which is chosen from early signs of reflected ions in the upstream (edge of the foot) to the last encounter of the reflection point at the downstream side (slightly beyond the magnetic overshoot). This avoids identifying features that are not associated with the shock.
    \item Transform the filtered, reduced distribution into a contour map to identify 3D minima and maxima.
    \item Filter contours based on the following conditions:
    a) Remove open contours, b) remove contours that correspond to maxima (when at least 60\% of points are higher than the contour level), c) remove contours that are too narrow (less than 0.2 seconds), c) remove contours that are too wide (greater than 10 seconds), d) keep contours where the geometric center is located inside the boundary (features that do not satisfy this have complex geometric features and are not holes - such as an L shape), and finally e) require that a contour is not isolated so that contours share at least two center points from all identified contours, in other words, a similarly shaped and nearby closed contour was identified.
    \item PSHs are then identified based on the contours with the maximum number of center points inside. 
    \item The width, $\Delta t_w$, and height, $\Delta V_w$, of PSHs are defined as the maximum width and height of the contour.
\end{enumerate}

The white contours in Figure \ref{fig:PSH_ID} show examples of the above PSH identification procedure for three separate shocks of varying PSH numbers. One PSH was detected for the shock on 2015-10-17, whereas 11 PSHs were found for the shock on 2016-01-06; no PSHs were found for the last example on 2016-11-10.

From limited events such as these, it is far from obvious how shock parameters are connected to the number of PSHs and their properties. Thus, this figure exemplifies the motivation of this study to take a statistical approach to this problem. The following section will investigate the relationship between PSHs and fundamental shock parameters.

\section{Data set}
This study employs crossings of Earth's bow shock made by the Magnetospheric Multiscale (MMS) mission between October 2015 and December 2020. 
The shock crossings in this study are taken from the MMS shock database compiled by \citeA{lalti2022database}, which contains 2797 bow shock crossings. This database utilizes the results of \citeA{Olshevsky2021Auto}, which employs a convolutional neural network to classify MMS data according to the attributes of the 3D ion distribution function. The MMS data is classified as the solar wind, magnetosheath, foreshock, and magnetosphere regions. Bow shock crossings are determined from the classification shift from the foreshock-magnetosheath, solar wind-magnetosheath, and \textit{vice versa}. For each shock, the database provides key shock parameters such as $M_A$, $\mathbf{\hat n}$, and $\theta_{Bn}$, as well as upstream solar wind properties such as the magnetic field vector $B_u$ and solar wind bulk velocity $V_{u}$. For all the shocks, $\mathbf{\hat n}$ is calculated from the Farris bow shock model \cite{farris1991thickness}, and upstream parameters are retrieved from the OMNI service \cite{King2005Sol}. We direct readers to \citeA{lalti2022database} for a more detailed database description. As the database did not include any shocks with $M_A<3$, we complement the dataset by five low-Mach number, $1.4 \leq M_A \leq 1.9$, bow shock crossings observed during a CME encounter \cite{graham2024_ion_dynamics_across,graham2025_Structure_and_Kinetic}. 

For this study, we extract quasi-perpendicular shocks from the database satisfying $\theta_{Bn}>45^{\circ}$. We also require burst mode data, which provides ion VDF measurements with 150~ms cadence, to resolve the PSHs. We select the shock transitions showing clear solar wind/magnetosheath regions. We identify 521 shock crossings for the study. Figures \ref{StatsFig1}a and \ref{StatsFig1}b show the statistical distributions of $M_A$ and $\theta_{Bn}$ for the shocks used in this study. Most shocks have $M_A \sim 5-10$; however, there is a tail with $M_A$ exceeding 20. The shock geometry is quite evenly distributed between $45^{\circ}-90^{\circ}$, with slightly more shocks between $75^{\circ}-90^{\circ}$. We note that the five low-$M_A$ shocks ($M_A< 2)$ have $\theta_{Bn} >65^{\circ}$. 

\section{Statistical Results}
\begin{figure*}[htbp!]
\begin{center}
\includegraphics[width=120mm, height=160mm]{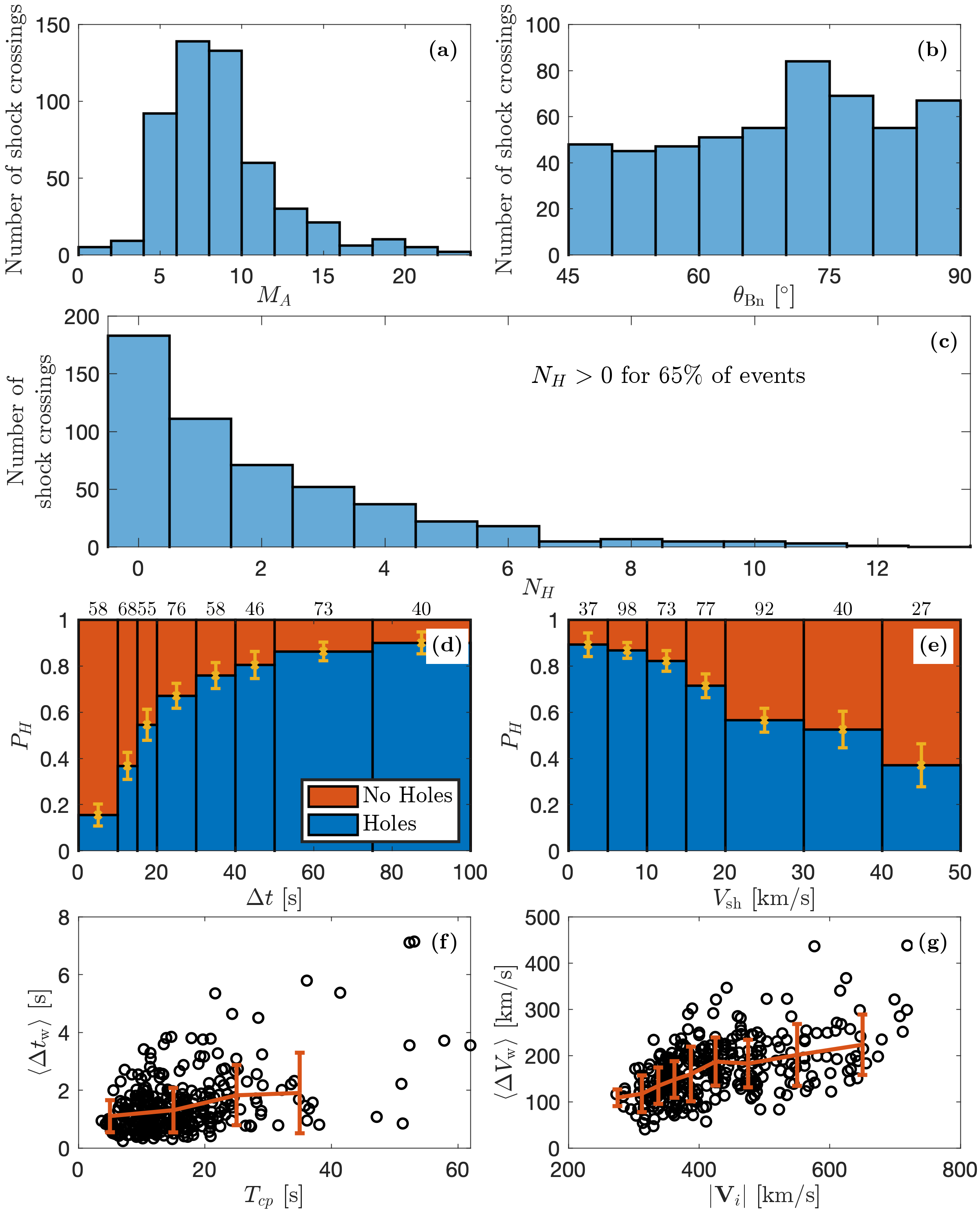}
\caption{Statistics of PSHs of the 521 shock crossings. (a) and (b) Histograms of the number of shock crossings versus $M_A$ and $\theta_{Bn}$. (c) Histogram of the number PSHs observed for each shock. The red line is the cumulative sum. (d) and (e) $P_H$ versus the shock crossing time $\Delta t$ and the shock speed in the spacecraft frame $V_{sh}$. The numbers above each bin indicate the number of shocks in each bin. (f) and (g) Scatter plots of the velocity width $\Delta V_w$ versus upstream $V_i$ and temporal width $\Delta t_w$ versus $T_{cp}$. The red lines show the median and standard deviations as a function of $\Delta V_w$ and $T_{cp}$. }
\label{StatsFig1}
\end{center}
\end{figure*}

We now investigate the statistical occurrence of the ion holes and their properties. Figure \ref{StatsFig1}c shows the histogram of the number of ion holes $N_H$ for each of the shock crossings. We find that $N_H \geq 1$ is observed at $65$~\% of the shock crossings. We see that shocks with PSHs$>$6 are rare, and thus the shock that was shown in Figure \ref{fig:PSH_ID} with 11 PSHs is not a typical event. We note that the PSH occurrence is not dependent on the location at the bow shock (not shown), i.e., no dependence on flank versus sub-solar point observations.

To further understand this result, Figures \ref{StatsFig1}d and \ref{StatsFig1}e show $P_H$, defined as the fraction of bow shock crossings with one or more PSHs to the total number of bow shocks within a given parameter range, for the shock crossing time $\Delta t$ and the shock speed $V_{sh}$. The shock speed is determined from the shock crossing time combined with an estimate of the shock foot width in the method described by \citeA{gosling1985specularly}. This method is the most suitable in these cases since the small spacecraft separation of MMS typically does not allow for four-spacecraft timing to determine the shock speed. Figure \ref{StatsFig1}d shows that the probability of PSH detection increases as $\Delta t$ increases. This is reinforced by Figure \ref{StatsFig1}e, which shows that the probability of PSH detection decreases as $V_{sh}$ increases. In other words, shocks moving slowly in the spacecraft frame provide more time for the spacecraft to observe PSHs. These results show that the observation of PSHs depends strongly on the observed shock speed, or equivalently the shock crossing time, and that the identification of ion holes underestimates the probability of rippling occurring at the bow shock. 

Figures \ref{StatsFig1}f and \ref{StatsFig1}g show the dimensions of the PSHs that were identified. The dimensions correspond to the scale of the PSHs in time, $\Delta t_w$, and normal velocity, $\Delta V_w$. 
Figure \ref{StatsFig1}f shows some indication of a dependency on the ion gyroperiod, $T_{cp} = 2 \pi m_p / e B$, especially since the spread increases for larger $T_{cp}$. Still, this dependency is unclear and deserves additional investigation. Figure \ref{StatsFig1}g shows that statistically, the width of the ion holes in $v_n$ increases with solar wind speed $V_i$. This implies that faster solar wind speeds will cause a stronger reflected ion population; since PSHs are measured between the incident and reflected ion populations, the vertical dimension will increase. We note that $\Delta t_w$ and $\Delta V_w$ are defined by the last closed contour (see section \ref{sec:method}) and thus systematically smaller than the actual PSH size; however, this does not affect the scaling with $T_{cp}$ and $V_i$.

\begin{figure*}[htbp!]
\begin{center}
\includegraphics[width=140mm, height=160mm]{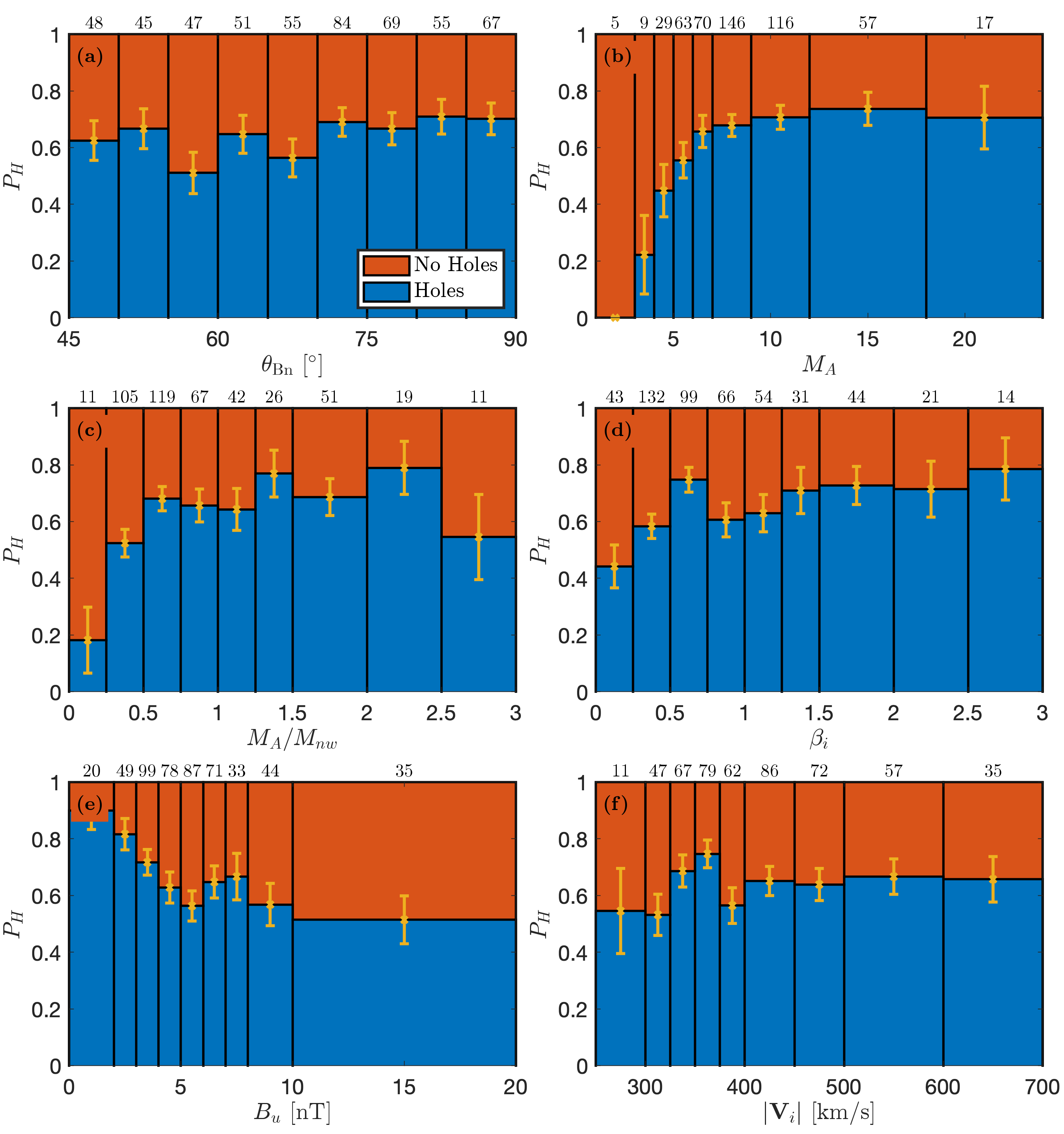}
\caption{Dependence of ion hole observations versus shock parameters. (a) $P_H$ versus $\theta_{Bn}$, (b) $P_H$ versus $M_A$, (c) $P_H$ versus $M_A/M_{nw}$, (d) $P_H$ versus upstream $\beta_i$, (e) $P_H$ versus upstream magnetic field strength $B_u$, and (f) $P_H$ versus upstream $V_i$. The numbers above each bin indicate the number of shock crossings in each bin.}
\label{StatsFig2}
\end{center}
\end{figure*}

Figure \ref{StatsFig2} shows the dependence of $P_H$ on $\theta_{Bn}$, $M_A$, $M_A/M_{nw}$, upstream $\beta_i$, $B_u$, and $V_i$. Here, $M_{nw} = |\cos \theta_{Bn}| (2 m_e/m_i)^{-\frac{1}{2}}$ is the nonlinear critical whistler Mach number \cite{krasnoselskikh2002nonstationarity}. From Figure \ref{StatsFig2}a, we find no clear dependence of $P_H$ on $\theta_{Bn}$. Figure \ref{StatsFig2}b shows that $P_H$ increases as $M_A$. For $M_A < 3$, no PSHs are observed, although only five shocks are considered. For these shocks, ion reflection was small or was not observed, and the downstream regions were quite laminar \cite{graham2025_Structure_and_Kinetic}. For $3 \lesssim M_A \lesssim 7$, $P_H$ sharply increases with $M_A$. This range of $M_A$ corresponds to the transition from subcritical to supercritical shocks, where strong proton reflection is expected. For $M_A \gtrsim 8$, $P_H$ remains relatively constant, meaning further increases in $M_A$ do not increase the likelihood of observing PSHs. 



Figure \ref{StatsFig2}c shows the dependence of $P_H$ on $M_A/M_{nw}$. We find that $P_H$ increases weakly with $M_A/M_{nw}$. For $M_A/M_{nw} > 1$, reformation is expected to occur due to the gradient catastrophe mechanism \cite{krasnoselskikh2002nonstationarity, dimmock2019direct}. However, we find that $P_H > 0.5$ for $M_A/M_{nw} < 1$, meaning that the gradient catastrophe mechanism is not required to produce the phase space holes. The slight increase in $P_H$ versus $M_A/M_{nw}$ may result from $P_H$ increasing with $M_A$.


We investigate $P_H$ dependence of upstream plasma ion beta, $\beta_i$ (Figure \ref{StatsFig2}d). 
We find that the dependence of $P_H$ on $\beta_i$ is weak, with $P_H$ slightly increasing with $\beta_i$. 

Figure \ref{StatsFig2}e shows $P_H$ versus $B_u$. We find that $P_H$ decreases as $B_u$ increases. Two explanations for this are: (1) As $B_u$ increases, the proton gyroradius decreases, and the proton cyclotron frequency increases, decreasing the probability of detecting ion holes. (2) As $B_u$ increases, $M_A$ will decrease for nominal solar wind speeds, reducing $P_H$. This dependence on $B_u$ may account for the dependence of $P_H$ on $\beta_i$. 

Figure \ref{StatsFig2}f shows that $P_H$ does not depend on the upstream speed $V_i$. This suggests that the observation of ion holes does not strongly depend on the observed $\Delta V_w$. Since the solar wind at 1 AU is characterized by $V_i/v_p \gg 1$ for the range of observed $V_i$, where $v_p$ is the proton thermal speed, the observation of ion holes likely does not depend strongly on $V_i$.


\section{Discussion}
We have investigated the occurrence rate of ion holes at Earth's quasi-perpendicular bow shock. We find that ion holes are detected at 65\% of the 521 bow shock crossings analyzed, and $P_H$ increases as $V_{sh}$ decreases, or equivalently, as the shock crossing time increases. We estimate the occurrence rate of ion holes at the quasi-perpendicular bow shock to be $\sim$90\% and represent a characteristic feature of Earth's quasi-perpendicular bow shock, based on the occurrence rates for the slowest moving shocks in the spacecraft frame, see Figure \ref{StatsFig1}. However, ion holes may not be observable for fast shock crossings, which has implications for understanding the in situ observations of fast-moving shocks, such as interplanetary shocks, where ion reflection can occur.

We find that the $P_H$ depends strongly on $M_A$, while $P_H$ is either weakly dependent or shows no dependence on other shock parameters, such as $\theta_{Bn}$, $M_A/M_{nw}$, $V_i$, and $\beta_i$. In particular, $P_H = 0$ for $M_A < 3$ (although only five shocks satisfied this criterion), while for $3 \lesssim M_A \lesssim 7$, $P_H$ sharply increases. This coincides with the transition from a subcritical to a supercritical bow shock. 

Previous studies \cite{Johlander2016rippled} have shown that ion holes are strong evidence of shock rippling. From these studies, it has not been possible to determine if such crossings are rare or a standard feature. Since ion holes are a signature of shock non-stationarity in the form of rippling, this study sheds light on this open question. A surprising result is the consistent detection of ion holes, meaning that non-stationarity of the quasi-perpendicular bow shock, in the form of rippling of the bow shock surface, could be a characteristic feature.

The ripples can be generated by a combination of shock reformation and the AIC instability \cite{Winske1988Mag,burgess2016}. Even though the threshold $T_{\perp}/T_{||}$ of the AIC instability decreases with the increase of local $\beta_i$ in the foot \cite{wu1984}, the anisotropy $T_{\perp}/T_{||}$ produced by the ion reflection is already high, so the threshold should be generally satisfied. Also, local $\beta_i$ in the foot will largely be determined by the ion reflection and have a weak relation to the upstream $\beta_i$. This might explain the weak observed dependence of the PSH occurrence on upstream $\beta_i$. However, this weak dependence can also be attributed to the ripple scales, which are determined by the gyroradius of the shock-reflected ions, becoming shorter with a decrease in upstream $\beta_i$, thus making it less likely to observe a PSH. 

In addition to the ripples, for $M_A$ above the whistler critical Mach number, $M_w$, the PSHs can also be related to non-linear standing whistler \cite{scholer2007_whistler_waves}. Separating such PSHs from ripples statistically is challenging, as this would require investigation of the associated $B_n$ oscillations, which are expected in the case of ripples \cite{burgess2006_interpreting_multipoint}. Such $B_n$ oscillations associated with PSHs were investigated in a small number of case studies \cite{Johlander2016rippled, gingell2017_MMS_bservations, Johlander2018}.

At higher Mach numbers, particularly for $M_A/M_{nw} > 1$, the gradient catastrophe mechanism can also generate reformation, causing similar PSH. However, PSHs are common even for $M_A/M_{nw} < 1$, corresponding generally to shocks with $\theta_{Bn}<75^{\circ}$. Thus, suggesting that the mechanism is not required to generate the holes and that other mechanisms are operating. 

These results show that Earth's quasi-perpendicular bow shock is typically non-stationary. This non-stationarity is due to the super-critical bow shock with strong ion reflection. The non-stationarity causes large oscillations in the local ramp speed \cite{johlander2023} and electric and magnetic fields of the ramp \cite{khotyaintsev2024ionreflectionrippledperpendicular}, and thus interpreting such structures as stationary can lead to erroneous conclusions. For very fast-moving super-critical shocks, such as interplanetary shocks \cite{dimmock2023,trotta2024}, PSHs are unlikely to be observed since the shock speed (along $\hat n$) greatly exceeds the speed of the ripples that would be mostly tangential to $\hat n$. However, ripples are likely present and also play a key role in shock dynamics. As a result, we can conclude that the observed structure of the ramp and foot is likely a transient feature.


\section{Conclusions}
We have conducted the first statistical study of the non-stationarity of the quasi-perpendicular bow shock and its association with shock parameters. The identification of non-stationarity is based on the detection of ion PSHs. From this study, we draw the following conclusions:

\begin{enumerate}
\item PSHs are common and were observed at 65\% of the examined shocks. However, the true occurrence is likely higher.
\item The number of observed PSHs at a given shock is generally low ($<$6), and highly rippled shocks (10 or more) are rare.
\item The observed shock speed and crossing time play a fundamental role in the ability of a spacecraft to observe PSHs since the number of PSHs was significantly larger for slower and longer shock crossings.
\item The probability of observing at least one PSH sharply increases with $M_A$ for $3 \lesssim M_A \lesssim 7$, suggesting supercritical shocks are crucial for ion hole formation.
\end{enumerate}

Overall, we find that the quasi-perpendicular bow shock is typically non-stationary (rippled) for nominal conditions ($M_A > 3$). This means that the shock structure is highly varying at ion gyro-period and spatial scales, which must be considered when analyzing such shocks, e.g., when estimating scales of shock structures \cite{johlander2023}.

\section*{Data Availability Statement}
MMS data are available at \url{https://lasp.colorado.edu/mms/sdc/public} and \url{https://spdf.gsfc.nasa.gov/pub/data/mms/}. We use burst mode magnetic field data from FGM \cite{mms1fgmbrst}. We use burst mode ion distributions \cite{mms1fpidisdistbrst} and moments \cite{mms1fpidismomsbrst} from FPI. The data analysis was performed using the irfu-matlab \cite{khotyaintsev_2024_11550091}.

\acknowledgments
This research was made possible with
the data and efforts of the team of the Magnetospheric Multiscale mission. This work is supported by the Swedish Research Council Grant 2018-05514, the European Union's Horizon 2020 research and innovation program under grant agreement number 101004131 (SHARP), and the Swedish National Space Agency.

\end{document}